\documentclass[twocolumn,pra,amsmath,amssymb,showpacs,nofootinbib]{revtex4}
\usepackage{graphicx}
\usepackage{dcolumn}
\usepackage{bm}
\usepackage{latexsym}
\usepackage{amstext}
\usepackage{amsxtra}

\newcommand{\kb}[0]{k_{\mathrm{B}}}
\begin{document}
\title{Evaporative Cooling of a Guided Rubidium Atomic Beam}
\author{T.~Lahaye, Z.~Wang, G.~Reinaudi, S.~P.~Rath, J.~Dalibard and D.~Gu\'ery-Odelin}
 \affiliation{Laboratoire Kastler Brossel$^{*}$, 24 rue Lhomond,
F-75231 Paris Cedex 05, France}
 \date{May 30, 2005}

\begin{abstract}
We report on our recent progress in the manipulation and cooling
of a magnetically guided, high flux beam of $^{87}{\rm Rb}$ atoms.
Typically $7\times 10^9$ atoms per second propagate in a magnetic
guide providing a transverse gradient of $800$~G/cm, with a
temperature $\sim550$~$\mu$K, at an initial velocity of 90~cm/s.
The atoms are subsequently slowed down to $\sim 60$~cm/s using an
upward slope. The relatively high collision rate (5~s$^{-1}$)
allows us to start forced evaporative cooling of the beam, leading
to a reduction of the beam temperature by a factor of~4, and a
ten-fold increase of the on-axis phase-space density.
\end{abstract}


 \maketitle

\section{Introduction}

Evaporative cooling of atomic clouds confined in magnetic or
optical traps has allowed for the realization of Bose-Einstein
condensation in dilute gases \cite{bec}. Coherent streams of
matter waves have been extracted from such condensates, leading to
the achievement of pulsed ``atom
lasers''\cite{laser1,laser2,laser3,laser4}. Nevertheless, the very
low duty cycles and mean fluxes of the atom laser experiments
realized to date are crucial inconveniences in view of practical
applications, such as metrology or nanolithography experiments.
Such applications would indeed require the equivalent, for matter
waves, of a high flux, continuous-wave laser.

A possible way to realize such a cw atom laser consists in
outcoupling atoms from a trap periodically replenished with new
condensates~\cite{ketterle}. Another way to achieve this goal has
been studied theoretically in \cite{mandonnet} and transposes the
technique of evaporative cooling to an \emph{atomic beam}. A high
flux, slow and cold atomic beam, transversely confined by a
magnetic guide, can be further cooled by applying an
energy-selective ``knife'' which removes atoms having a transverse
energy above the average. As the remaining particles propagate
further downstream and rethermalize through elastic collisions,
the beam temperature decreases. By lowering the knife as the beam
gets colder, forced evaporative cooling takes place, and the
phase-space density of the beam increases.

The challenging prerequisite in order to apply such an evaporative
cooling technique is to realize a guided atomic beam in the
collisional regime, as demonstrated for the first time in
\cite{PRLRb2}. Here, we report on several improvements of the
setup allowing us to achieve a ten-fold increase of the beam
phase-space density by evaporative cooling. An upward slope is
implemented along the first 1.7~m of the guide, leading to a
significant decrease of the beam velocity. This, in combination
with an optimization of the pulsed injection of atoms into the
guide, yields a significant increase in the collision rate. We
have also developed a new detection scheme using absorption
spectroscopy on an open transition, which improves the accuracy of
our temperature measurements.

The article is organized as follows. In section \ref{Par:Setup} we
describe the experimental setup, with a special emphasis on the
implementation of the tilted guide which allows us to decrease the
beam mean velocity. In section \ref{Par:Majorana} we show the
occurrence of non-adiabatic spin-flip losses due to the magnetic
field vanishing on the guide axis. We explain how this can be
circumvented by applying a longitudinal bias field. We also
discuss the necessary modification of the analysis made
in~\cite{epjd} of the temperature measurements deduced from RF
spectroscopy on the atomic beam. Finally, in the last section, we
report on what is, to our knowledge, the first significant gain in
the phase-space density of a guided atomic beam, obtained by
continuous evaporative cooling.

\section{Experimental setup}
\label{Par:Setup}

\subsection{Magnetic guide}

\begin{figure*}[t]
\includegraphics[width=17cm]{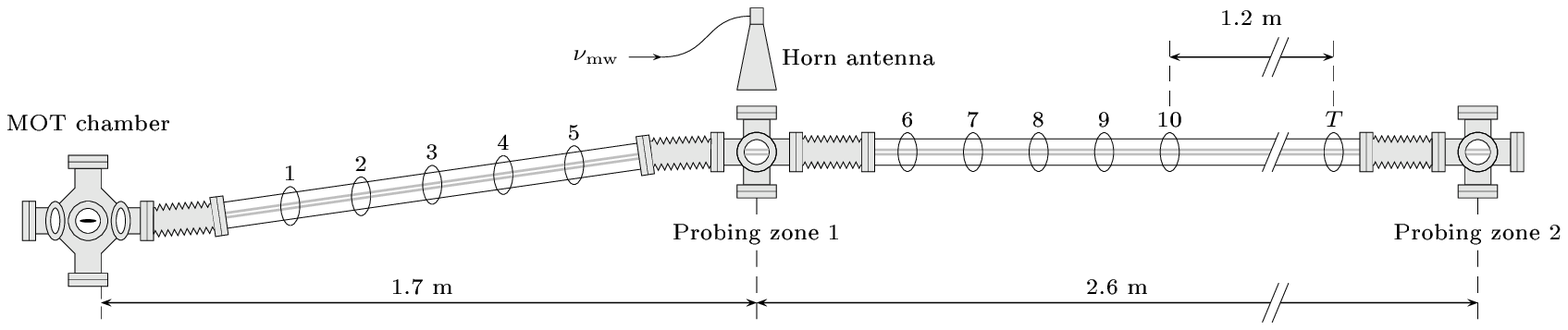}
\caption{Simplified sketch (not to scale) of the magnetic guide.
The first 1.7~m of the guide is slightly tilted upwards by
$h_0=22\pm1$~mm, while the remaining part is horizontal within
$\pm 1$~mm. The guide sits inside glass tubes so that
radio-frequency (RF) waves can be shone on the atomic beam. Two
probe beams, one just after the slope, the other close to the end
of the guide, can be used to detect the atoms. The beam in the
\emph{Probing zone~1} can also be used as a marking beam for
longitudinal time of flight measurements (see text). The ellipses
numbered from 1 to 10 represent the RF antennas used for the
evaporative cooling of the beam; the one labelled $T$ represents
the RF antenna used to measure the beam temperature.}
\label{Fig:Setup}
\end{figure*}

The magnetic guiding
\cite{guide0,guide00,guide1,guide2,guide3,guide4} of our atomic
beam is ensured by a two-dimensional quadrupole guide. The guide
axis being chosen as the $z$-axis, the magnetic field reads
$(bx,-by,0)$, where $b$ is the magnetic field gradient. The field
increases linearly with the distance from the axis, where it
vanishes, which leads to non-adiabatic spin-flip losses (see
section~\ref{Par:Majorana}). Contrary to what happens with a
spherical quadrupole field, this loss mechanism can be
circumvented very simply by adding a longitudinal bias field
$B_0$. The magnetic field is then non-zero everywhere. Atoms
prepared in a low-field seeking state with a magnetic dipole
moment $\mu$ thus experience a semi-linear cylindrically symmetric
potential given by:
\begin{equation}
U(r)=\mu\sqrt{B_0^2+b^2r^2}\,, \label{Eq:Potential}
\end{equation}
where $r=(x^2+y^2)^{1/2}$ is the radial distance from the axis. In
our experiments, we use $^{87}$Rb atoms polarized in the
$|F=1,\,m_F=-1\rangle$ state, which has a magnetic dipole moment
$\mu=\mu_{\rm B}/2$, where $\mu_{\rm B}$ is the Bohr magneton.

We denote by $\Phi$ the flux of the atomic beam propagating in the
guide, $\bar v$ its mean velocity and $T$ its temperature.
Depending on the value of the dimensionless parameter
$\alpha\equiv \mu B_0/(k_{\rm B}T)$, the potential felt by the
atoms is approximately linear (if $\alpha\ll 1$) or approximately
harmonic ($\alpha\gg 1$). For arbitrary $\alpha$, exact analytical
expressions for the elastic collision rate $\gamma$ within the
beam, as well as for the on-axis phase space density $\rho$, can
be derived for a semi-linear potential:
\begin{equation}
\gamma=\frac{\sigma}{2\pi^{3/2}}\,\frac{1+2\alpha}{(1+\alpha)^2}\,\frac{\Phi}{\bar
v}\,\left(\frac{\mu b}{\kb T}\right)^2\sqrt{\frac{\kb T}{m}}\,,
\label{Eq:Cr}
\end{equation}
\begin{equation}
\rho= \frac{1}{2\pi}\,\frac{1}{1+\alpha}\,\frac{\Phi}{\bar
v}\,\left(\frac{\mu b}{\kb T}\right)^2\,\frac{h^3}{(2\pi m \kb
T)^{3/2}}\,, \label{Eq:Psd}
\end{equation}
where $m$ is the atomic mass. In Eq.~(\ref{Eq:Cr}), the elastic
collision cross-section $\sigma$ is assumed to be purely $s$-wave
(and thus isotropic) and velocity-independent ($\sigma=7.6\times
10^{-16}$~m$^2$). Both assumptions are known to be valid only if
$T<100$~$\mu$K for $^{87}$Rb. However the temperatures
investigated in this article range from 150~$\mu$K to 600~$\mu$K.
For those temperatures, the $d$-wave contribution has to be taken
into account in the calculation of both the collision rate and the
thermalization rate. The inclusion of the $d$-wave contribution in
the estimation of the thermalization rate is not a straightforward
task. One problem is the anisotropy of the $d$-wave differential
cross-section, which reduces the efficiency of the thermalization
for a given number of collisions. Another difficulty lies in
interferences between $s$ and $d$-waves contributions. For the
particular case of $^{87}$Rb, the $d$-wave cross section exhibits
a resonance~\cite{servaas} which increases the collision rate
around 300~$\mu$K. This compensates partially for the reduced
efficiency in rethermalization of $d$-wave scattering, resulting
in a nearly constant thermalization rate~\cite{dgo}. Therefore, in
the following, we use the expression (\ref{Eq:Cr}) as a guideline
to estimate the number of collisions.

We now turn to the experimental realization of such a
guide~\cite{cren}. Currents up to $I=400$~A are running in four
4.5~m long parallel copper tubes sitting at the vertices of a
square of side $a=8$~mm. The currents change sign between adjacent
tubes. A ring-shaped, hollow, non-magnetic stainless-steel piece,
brazed on the copper tubes, is used at the entrance of the guide
in order to recombine the currents and the cooling water used to
dissipate the $\sim 5$~kW generated by Ohmic heating. The copper
tubes are held together using ceramic pieces every 30~cm, and the
guide rests on other ceramic pieces which themselves rest on the
inner side of the vacuum chamber. For this configuration of
currents, the magnetic field gradient is $b=4\mu_0 I/(\pi a^2)$,
and can therefore reach a maximum value of $1$~kG/cm for
$I=400$~A. The magnetic field gradient is actually smaller by a
factor of three at the entrance of the guide (where $a=14$~mm) and
reaches its asymptotic value only after $40$~cm as the distance
$a$ is progressively reduced. The goal of this tapered section is
to compress adiabatically the beam during its entrance into the
guide, in order to maximize the collision rate $\gamma$.

The whole guide sits in an ultra-high vacuum chamber (see
Fig.~\ref{Fig:Setup}) consisting essentially of two glass tubes
allowing us to use RF antennas located outside of the vacuum
chamber, since metallic tubes would screen the oscillating
magnetic field. Two 25~l/s ion pumps located at $z=1.7$~m and
$z=4.3$~m from the guide entrance ensure that the pressure remains
around $10^{-11}$~Torr at the pumps locations; however, due to the
poor conductance of the glass tubes, the effective lifetime in the
second part of the guide, measured by monitoring the flux at
$z=1.7$~m and $z=4.3$~m for various beam velocities, is limited to
$\sim 15$~s. We find experimentally that cooling the vacuum
chamber to below $-50^\circ$~C increases the flux measured at the
end of the guide by 25\% to 50\%. For this purpose, we use a flow
of cold nitrogen gas emanating from a liquid nitrogen dewar. This
increase is a clear indication that the main source of background
gas is the outgassing from the glass walls. In addition, we have
found that the repeated, fast switching-off of the guide current
leads to a degradation of the vacuum, presumably by causing
micro-leaks to appear at the level of the brazing between the
copper tubes and the recombination piece at the entrance of the
guide. For this reason, we do not turn off the magnetic field of
the guide when we measure the atomic beam flux, and we
consequently developed a flux measurement technique (see
\S~\ref{Par:Probe}) which is almost insensitive to the presence of
the magnetic field.

\subsection{Loading of the guide}
\label{Par:Pulsed}

In order to produce a high-flux, low velocity, cold atomic beam
propagating in the guide, we use a pulsed injection scheme as
described in detail in Ref.~\cite{PRLRb2}. We recall here briefly
the various steps involved in the 180~ms sequence used to inject a
``packet'' of atoms into the guide.

A slow ($\sim20$~m/s) atomic beam, with a flux on the order of
$2\times 10^{11}$ atoms/s, is produced by decelerating, in a
Zeeman slower (not shown in Fig.~\ref{Fig:Setup}), an atomic beam
effusing from an oven at $140^\circ$~C. It is used to feed a
three-dimensional magneto-optical trap (MOT). The MOT magnetic
field reads $(b_x x,\,b_y y,\,b_z z)$ with
$(b_x,b_y,b_z)=(5.4,-6,0.6)$~G/cm. This anisotropy in the field
gradients results in an elongated shape of the cloud. After 100~ms
of loading, the MOT contains about $1.8\times10^{9}$ atoms, and
has a transverse rms radius of about $\Delta r=1$~mm and a length
of 35~mm along~$z$. A 3~ms launching phase occurs, in which the
MOT magnetic fields are switched off, and a small frequency offset
$\delta\nu$ is applied between the front and rear pairs of the MOT
beams lying in the horizontal plane. In this \emph{moving
molasses} configuration~\cite{launch}, the cooling occurs in a
frame moving with a velocity $v_{\rm inj}\propto\delta\nu$. This
way, the atom cloud is set in motion towards the guide entrance,
with an injection velocity $v_{\rm inj}$ adjustable between 30 and
250~cm/s. In the last 1.5~ms of this phase, the atoms are further
cooled by increasing the laser detuning from the cooling
transition to $-72$~MHz. This results in a final cloud temperature
around 40~$\mu$K, hence a rms velocity $\Delta v = 6$~cm/s.

A 700~$\mu$s optical pumping stage is used to pump the atoms into
the $|F=1,\,m_F=-1\rangle$ ground state, with an efficiency of
$\sim 70$\%. A two-dimensional quadrupole magnetic ``pre-guide''
is then turned on for $80$~ms, with a transverse gradient in the
$x$-$y$ plane of 85~G/cm. It prevents the atom cloud from
expanding transversally and from falling due to gravity while it
propagates towards the guide entrance. In order to ``mode-match''
the pre-guide stiffness with the cloud size and temperature, one
can in principle superimpose a longitudinal bias field $B_1$. This
makes the pre-guide harmonic, with an angular frequency
$\omega_1\propto B_1^{-1/2}$. For the parameters of our atomic
cloud, the field needed to obtain $\omega_1=\Delta v/\Delta r$ is
$B_1=60$~G. Nevertheless, we find experimentally that applying
such a bias field does not lead to appreciable changes in the beam
temperature and flux. Moreover, the ``center matching''
(\emph{i.e.}, the overlapping of the atom cloud axis with the
guide axis~$z$), which is a crucial step in order to achieve a low
temperature of the guided beam, is greatly simplified in the
absence of bias field as there is no gravitational sag, whereas a
bias field $B_1=60$~G leads to a 3~mm sag.

The atomic packets containing $1.3\times10^9$ particles, and
injected with a rate of (180~ms)$^{-1}$, expand longitudinally
(due to the dispersion of longitudinal velocities) while they
enter into the guide, and finally overlap, resulting, after a
propagation over $50$~cm, in a continuous beam with a flux
$\Phi=7\times10^{9}$~atoms/s and a mean velocity $v_{\rm inj}$.

The timings cited here (100~ms of MOT loading and 80~ms of
pre-guiding) are optimized for $v_{\rm inj}=90$~cm/s. For other
injection velocities, these values have to be adjusted
accordingly. However, it is important to point out that the pulsed
injection scheme used here prevents us from injecting the beam at
arbitrary low velocities. Indeed, the pre-guiding time $\tau_{\rm
pg}$ must be long enough to let the packet propagate over a length
$\ell$ of typically 10~cm so that the preparation of the next
packet does not affect the previous one significantly. Therefore,
for injection velocities much lower than 90~cm/s, the pre-guiding
time $\tau_{\rm pg}\sim\ell/v_{\rm inj}$ would become
prohibitively long. This would lead to a low repetition rate
affecting the value of the coupled flux~$\Phi$, since the number
of atoms that one can accommodate in the MOT quickly saturates.

\subsection{Detection of the atomic beam}
\label{Par:Probe}

Absorption spectroscopy is used to monitor the beam flux, using
two laser beams located in the middle and at the end of the guide
(Probing zones 1 and 2 on Fig.~\ref{Fig:Setup}).

We have investigated two detection schemes. The first one consists
in using probe beams resonant with the $|5^2{\rm
S}_{1/2},F=2\rangle\to|5^2{\rm P}_{3/2},F'=3\rangle$ transition,
with a small amount of repumping light used to first repump the
atoms from $|5^2{\rm S}_{1/2},F=1,m_F=-1\rangle$ to $|5^2{\rm
S}_{1/2},F=2\rangle$. This has the advantage of yielding a
continuous signal if the probe beam is locked on the transition.
This allows, for example, to monitor the time dependence of the
flux. The velocity measurements described in the next section are
realized this way.

However, the guide magnetic field is present during those
measurements. It leads to a large inhomogeneous broadening of the
resonance line, largely dependent on the field gradient $b$ and on
the probe beam polarization. We find that in practice, this
measurement method is not reliable if one wants to measure the
atomic beam flux independently of the transverse energy of the
atoms. For the temperature measurements through RF spectroscopy
(see below), where one is interested in such a measurement, this
method can thus lead to erroneous results.

We therefore developed another technique, consisting in monitoring
the absorption of the repumping beam alone. The atomic transition
is obviously an open one, and the absorption is therefore quite
small, on the 1\% level. Nevertheless, provided the light
intensity is high enough so that any atom, regardless of its
position or velocity, is addressed by the laser, the \emph{number}
of photons absorbed per second is proportional to the atomic flux.
In particular, for the transition $|5^2{\rm
S}_{1/2},F=1\rangle\to|5^2{\rm P}_{3/2},F'=2\rangle$ in which no
dark state exists (in contrast to what happens for $F'=0$ or
$F'=1$), each atom scatters on average two photons before being
pumped into the $|5^2{\rm S}_{1/2},F=2\rangle$ level. This is
therefore a convenient way to measure the absolute atomic flux in
the presence of the inhomogeneous magnetic field of the guide. We
find a good agreement between the flux deduced by this method and
the one inferred from the loading rate of the MOT.

In practice, for measurements of relative flux, we obtain the
better signal to noise ratio when we scan 70 times per second the
laser frequency across the $|5^2{\rm
S}_{1/2},F=1\rangle\to|5^2{\rm P}_{3/2},F'=0,1,2\rangle$
transitions, at a rate of about $600$~MHz/ms. For a beam velocity
of $\bar{v}=60$~cm/s, the atoms which are in the probing region
(of length 1~mm) for a given scan were thus located $\simeq 1$~cm
upstream during the previous scan and they were not affected by
the probe. We find that, in contrast to the previous detection
scheme, the spectra are remarkably insensitive to $b$ and to probe
polarization. This robustness of the flux signal thus leads to a
much more accurate temperature determination.

\subsection{Implementation of a slope to slow down the beam}

\begin{figure}
\includegraphics[width=7.3cm]{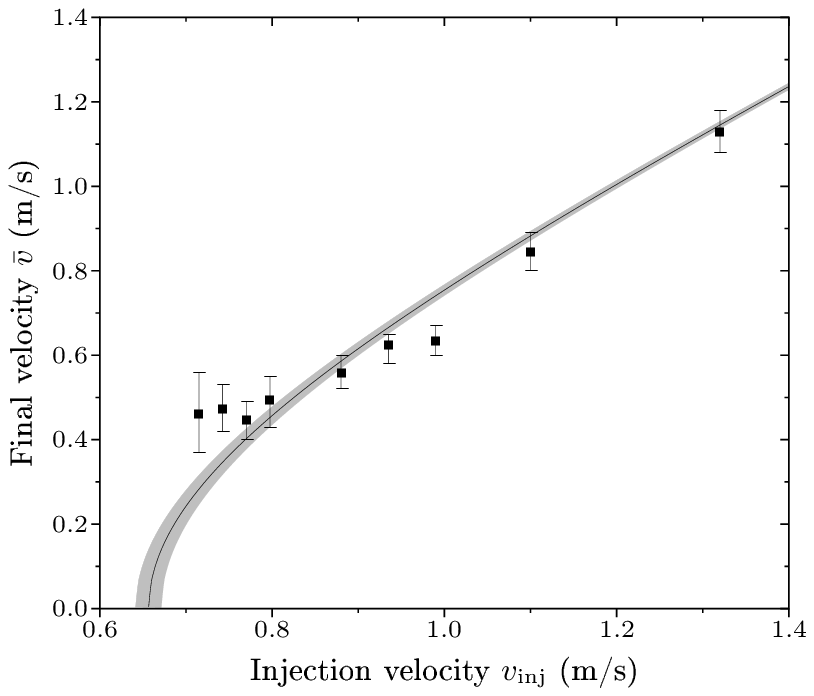}
\caption{Velocity $\bar{v}$ of the beam after the slope as a
function of the injection velocity. The magnetic gradient is
$b=800$~G/cm. The solid line represents $\sqrt{v_{\rm
inj}^2-2gh_0}$ with $h_0=22\pm 1$~mm (the shaded area represents
the uncertainty in $\bar{v}$ due to the uncertainty in $h_0$).}
\label{Fig:Slope}
\end{figure}

\begin{figure}
\includegraphics[width=7.3cm]{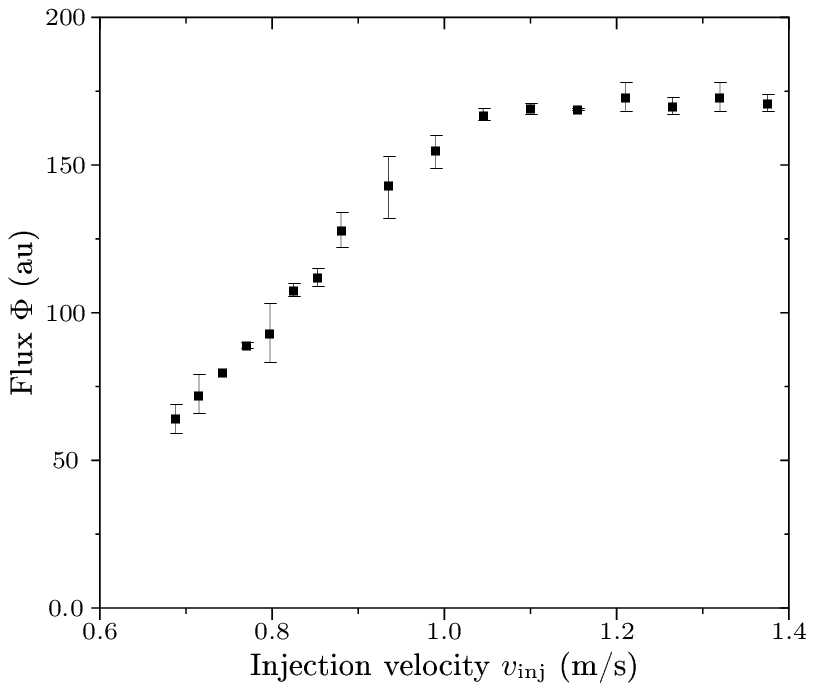}
\caption{Flux $\Phi$ measured at the end of the guide as a
function of the injection velocity $v_{\rm inj}$. The magnetic
guide gradient is $b=800$~G/cm. The timings of the launching
sequence are kept constant with the values of Section
\ref{Par:Pulsed}. One clearly sees that for low injection
velocities, the flux decreases very significantly.}
\label{Fig:FluxSlope}
\end{figure}

The number of collisions $N_{\rm col}$ experienced by an atom
during its propagation along the whole guide scales as
$1/\bar{v}^2$ where $\bar{v}$ is the beam mean velocity. Since the
value of $N_{\rm col}$ determines the potential gain in
phase-space density that one can achieve through evaporative
cooling~\cite{epjd}, one needs to minimize $\bar{v}$. But as
pointed out in \S~\ref{Par:Pulsed}, the injection velocity $v_{\rm
inj}$ cannot be arbitraryly low because of the pulsed character of
our loading process. Therefore, exploiting the relative mechanical
flexibility of the guide and the presence of bellows in the vacuum
system (see Fig.~\ref{Fig:Setup}), we implemented an upwards slope
on the first 1.7~m of the guide, the height of the guide rising by
$h_0=22\pm1$~mm. The remaining part of the guide is horizontal
within $\pm 1$~mm (see Fig. \ref{Fig:Setup}). The purpose of such
a slope is to decrease the mean velocity of the beam and thus
increase $N_{\rm col}$. In the absence of collisions, the product
of the mean velocity of the beam by the longitudinal velocity
dispersion remains constant in the slope as a result of
Liouville's theorem. In a collisional beam this effect yields a
slight increase of temperature by an amount $\delta T/T\sim
(2/7)\delta \bar{v}/\bar{v}$~\cite{lahaye}.

We have performed velocity measurements in the second part of the
guide with a longitudinal time-of-flight technique. The laser beam
in the \emph{Probing zone 1} on Fig. \ref{Fig:Setup}, located at
the position $z_1$, is scanned at a 200~Hz rate across the
transitions $|5^2{\rm S}_{1/2},F=1,m_F=-1\rangle\to |5^2{\rm
P}_{3/2},F',m_{F'}\rangle$ and is used as a marking beam. For a
power of 6~$\mu$W, typically 30\% of the atoms are removed from
the beam. At time $t=0$, we extinguish the marking beam, and we
monitor with the probe located in the \emph{Probing zone 2}
(locked on the cycling transition) the flux arriving at $z_2$. The
time-dependent signal is fitted according to
\begin{equation}
\Phi(t)=\Phi_0+\Delta\Phi\,{\rm
erf}\left(\frac{z_2-z_1-\bar{v}t}{t\sqrt{2\kb T/m}}\right)\,,
\end{equation}
where ${\rm erf}$ is the error function, with $T$ and $\bar{v}$ as
adjustable parameters.

Fig. \ref{Fig:Slope} represents the final velocity $\bar{v}$
measured this way, as a function of the injection velocity $v_{\rm
inj}$. One can see a clear reduction of the velocity, with
$\bar{v}$ close to the velocity  $\sqrt{v_{\rm inj}^2-2gh_0}$
expected for a single particle having an initial velocity $v_{\rm
inj}$ and ``climbing'' a height $h_0$ (solid line).

The use of a slope to slow down the beam is nevertheless limited
by the fact that one must stay in the supersonic regime: the
\emph{Mach number}, defined as $\mathcal{M}=\bar{v}/\Delta v$,
where $\Delta v$ is the longitudinal velocity dispersion, must
remain significantly larger than~$1$. If this is not true, many
particles would move backwards, thereby reducing the flux (see
Fig. \ref{Fig:FluxSlope}) and increasing the incoming particles
energy through collisions with a high relative momentum. In
practice, we find that for $v_{\rm inj}=90$~cm/s, the final
velocity $\bar{v}\simeq60$~cm/s (see Fig.~\ref{Fig:Slope}) is
still high enough for the beam to stay sufficiently supersonic
(for $T\sim 600$~$\mu$K, $\Delta v\sim 25$~cm/s). The net gain,
induced by the use of a slope, on the number of collisions $N_{\rm
col}$ is typically a factor of two for this choice of parameters.

\section{From linear to harmonic guiding potential}
\label{Par:Majorana}

\subsection{Spin-flip losses}

In a purely two-dimensional quadrupole guide, the magnetic field
vanishes along the guide axis $z$. Therefore, atoms having a low
angular momentum experience a rapid variation of the direction of
the magnetic field when they approach the axis, and can undergo a
non-adiabatic spin-flip transition to an untrapped state
\cite{cornell}. This loss mechanism is more pronounced if the beam
enters the collisional regime, since collisions redistribute
angular momentum between atoms and continuously lead to new
losses. Adding a longitudinal bias field to the two-dimensional
quadrupole leads to a semi-linear transverse trapping potential
(\ref{Eq:Potential}), in which the magnetic field never vanishes,
which can strongly suppress those Majorana losses.

\begin{figure}
\includegraphics[width=7.5cm]{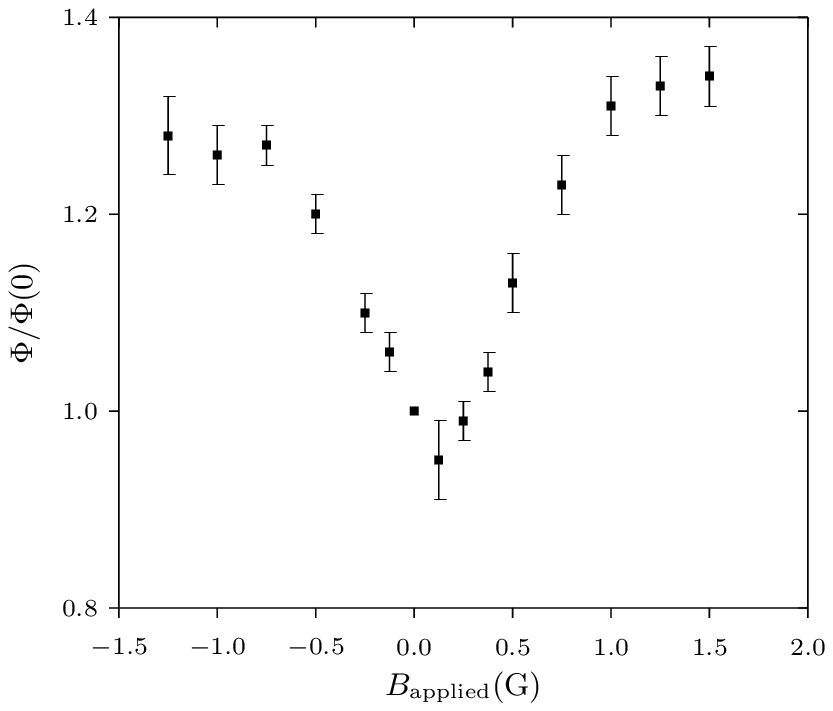}
\caption{Measured flux at the end of the guide (normalized to the
flux in the absence of an applied bias field) as a function of the
applied bias field $B_{\rm applied}$. The curve is not centered on
$0$ because of the presence of a $B_{\rm earth}\sim 120$~mG
component of the Earth magnetic field along the guide. The bias
field is $B_0=B_{\rm applied}-B_{\rm earth}$.}
\label{Fig:Majorana}
\end{figure}

To study the occurrence of such spin-flip losses in our beam, we
use a solenoid wound around the guide, with a diameter varying
from 30 to 50~cm and 20 turns per meter, yielding a longitudinal
bias field $B_0=0.25$~G/A along $z$. The resulting field is
uniform within $\pm10$\% along the whole guide length.

When we apply a sufficiently large bias field ($B_0>1$~G) the flux
in the guide is enhanced by as much as 30\% (see Fig.
\ref{Fig:Majorana}) with respect to its value for $B_0=0$. The
large width of the loss curve (1~G FWHM) indicates that the
observed losses are not, strictly speaking, due to Majorana
spin-flips, but are probably induced by high-frequency oscillating
fields (due to the noisy environment) having an amplitude of a few
tenths of gauss~\cite{majorana1,majorana2,majorana3}.

In practice, we operate the magnetic guide with a $B_0=1$~G bias
field. This changes slightly the shape of the transverse trapping
potential. If the beam temperature $T$ is large ($k_{\rm B}
T\gg\mu B_0$) the trap is still approximately linear; but for
lower temperatures the relevant part of the potential is harmonic.

\subsection{Temperature measurements in a semi-linear potential}

The exact shape of the potential needs to be taken into account in
the temperature measurements that we perform with RF
spectroscopy~\cite{PRLRb2}. For a linear potential, we extract the
beam temperature $T$ from the fraction $\varphi$ of atoms
remaining in the beam after evaporation by a RF antenna operating
at frequency $\nu_{\rm rf}$. If this frequency is small compared
to $k_{\rm B} T/h$, then very few atoms will fulfill the RF
resonance condition at some point in their trajectory and will be
evaporated. In a similar way, if $h \nu_{\rm rf}\gg k_{\rm B} T$,
the resonance condition is fulfilled on a radius large compared to
the thermal radial size of the beam, and almost no atoms are
evaporated. In between, the function $\varphi$ goes through a
minimum. The same feature remains qualitatively true for a
semi-linear potential, but the precise lineshape of this RF
spectrum depends on the value of the bias field.

We therefore developed, for an arbitrary bias field $B_0$,
semi-analytical calculations allowing us to extract the
temperature from the curve depicting the fraction
$\varphi(\nu_{\rm rf})$ of remaining atoms after the temperature
measurement antenna, as a function of its RF frequency $\nu_{\rm
rf}$. The method follows the one used in \cite{epjd}. The curve
$\varphi(\nu_{\rm rf})$ now depends on two dimensionless
parameters, $\eta=h(\nu_{\rm rf}-\nu_0)/(k_{\rm B} T)$ (where
$\nu_0$ is the frequency corresponding to the ``trap bottom'',
satisfying $h\nu_0=\mu B_0$) and $\alpha=\mu B_0/(k_{\rm B} T)$.
In the limit $\alpha\to 0$ one has a linear trap, leading to the
approximate formula
\begin{equation}
\varphi(\eta,0)\simeq1-1.65\,\eta^{1.13} {\rm e}^{-0.92\eta}\,,
\end{equation}
while for $\alpha\to\infty$ one recovers the exact formula for a
harmonic confinement:
\begin{equation}
\varphi(\eta,\infty)=1-\sqrt{\pi\eta}\,{\rm e}^{-\eta}\,.
\end{equation}
From our calculation, we find a good interpolation between those
two limits is provided by the following empirical formula:
\begin{equation}
\varphi_{\rm approx}(\eta,\alpha)\simeq1-1.7\,{\rm
e}^{-0.9\eta}\,\eta^{1.1-0.4\arctan(3.6\alpha)}\,.
\label{Eq:FitFunc}
\end{equation}
We checked the validity of the approximate formula
(\ref{Eq:FitFunc}) by generating, with a Monte-Carlo simulation,
ensembles of atoms at thermal equilibrium in the semi-linear
potential, simulating the curve $\varphi(\eta,\alpha)$ by applying
the evaporation criterion, and fitting this curve by Eq.
(\ref{Eq:FitFunc}). We find that the temperature obtained in this
way is reliable within $\pm5$\% for a range of variation of
$0.1\leqslant\alpha\leqslant 10$. The precision with which we can
infer temperature ratios is actually better. We have checked that,
for a given $B_0$, when two temperatures $T_1^{\rm sim}$ and
$T_2^{\rm sim}$ are used to obtain simulated RF spectra, the ratio
of the temperatures $T_1^{\rm fit}$ and $T_2^{\rm fit}$ obtained
by fitting those spectra with Eq. (\ref{Eq:FitFunc}) differs by
typically less than one percent from the actual ratio $T_1^{\rm
sim}/T_2^{\rm sim}$.

With this fitting function, we investigate the effect of spin-flip
losses on the beam temperature. In the absence of an applied
longitudinal field, we observe that the beam temperature is $\sim
8$\% higher than when one applies a bias field large enough to
avoid spin-flip losses. This can be easily understood since the
atoms lost by spin-flips have a low angular momentum, and on
average a small energy. Therefore their loss leads to a higher
beam temperature after rethermalization.

\section{Evaporative cooling of the beam}

It is important to emphasize that applying evaporative cooling on
an atomic beam is significantly more difficult than on a trapped
cloud of atoms: (i) the pulsed injection results in a longitudinal
dilution of the atom packets, and the resulting decrease in the
atomic density yields a low initial collision rate~$\gamma$, (ii)
the time allowed for evaporation is limited by the guide length
and by $\bar{v}$, and not only by the quality of the vacuum, and
(iii) the collision rate and phase-space density scale less
favorably in view of evaporative cooling for a two-dimensional
confinement than for a three-dimensional trap~\cite{epjd}. For all
those reasons, a significant gain in the phase-space density of a
magnetically guided beam had never been achieved to date.

In this section, after detailing the two evaporation methods that
we have developed, namely radio-frequency and microwave
evaporation, we report on the evaporative cooling of the atomic
beam using several evaporation zones in order to significantly
increase the phase-space density.

\subsection{Radio-frequency evaporation}

Evaporative cooling relying on a position-dependent transition to
an untrapped state can be easily accomplished using a two-photon
radio-frequency transition between the Zeeman sublevels
$|F=1,m_F=-1\rangle$ and $|F=1,m_F=1\rangle$. Atoms crossing a
cylinder of axis $z$, with a radius $R_{\rm evap}$ fulfilling
$U(R_{\rm evap})=h\nu_{\rm rf}$, where $h$ is Planck's constant
and $\nu_{\rm rf}$ the RF frequency, are evaporated. If this
radius is higher than the beam's thermal size, essentially
energetic atoms are removed, and after rethermalization, the beam
gets colder. For temperatures in the range of a few hundreds
$\mu$K, the corresponding frequencies are in the tens of MHz. The
RF antennas that we use consist of a single loop with a
50~$\Omega$ resistor in series, in which we couple a power of
$27$~dBm. This corresponds to an amplitude of the oscillating
magnetic field, at the level of the atomic beam, of 50~mG.

We can measure the spatial range over which an antenna is
efficient by switching on a RF antenna for a short time (a few
tenths of milliseconds) and monitoring the temporal width of the
dip appearing in the flux signal. We find that the typical range
is around 20~cm.

Nevertheless, for some specific frequencies (especially around
40~MHz) we observe that the efficiency range of the antenna
considerably increases, essentially reaching the whole guide
length. Those \emph{resonance} frequencies depend on the position
$z_{\rm ant}$ of the antenna on the guide. This fact, added to the
similarity between the guide length and the wavelengths
corresponding to the resonance frequencies, leads us to suppose
that the RF antennas can couple to the guide, in which they induce
currents, making it act as an antenna itself.

Although those resonances are well localized in the frequency
domain, and therefore do not prevent the use of RF evaporation for
the purpose of evaporative cooling, we tried to circumvent this
problem with evaporation in the microwave domain by inducing
transitions between different \emph{hyperfine} states.

\subsection{Microwave evaporation}

An electromagnetic wave resonant with the hyperfine transition
$|5^2{\rm S}_{1/2},\,F=1\rangle\to|5^2{\rm S}_{1/2},\,F=2\rangle$
at $\nu_{\rm hf}=6834.7$~GHz can also be used to selectively
remove atoms from the beam~\cite{hulet}. Indeed, a magnetic dipole
transition from the trapped state $|F=1,m_F=-1\rangle$ can only
excite atoms to the anti-trapped states $|F=2,m_F=-2\rangle$,
$|F=2,m_F=-1\rangle$, or to the untrapped state
$|F=2,m_F=0\rangle$. Contrary to the RF-case, this gives three
distinct evaporation radii. This leads actually to a better
efficiency of the evaporative cooling process compared to
single-radius evaporation, as was discussed in \cite{epjd} for the
case of evaporation with a continuum of radii spanning the range
$[R_{\rm evap},\infty)$.

\begin{figure}[t]
\includegraphics[width=7.5cm]{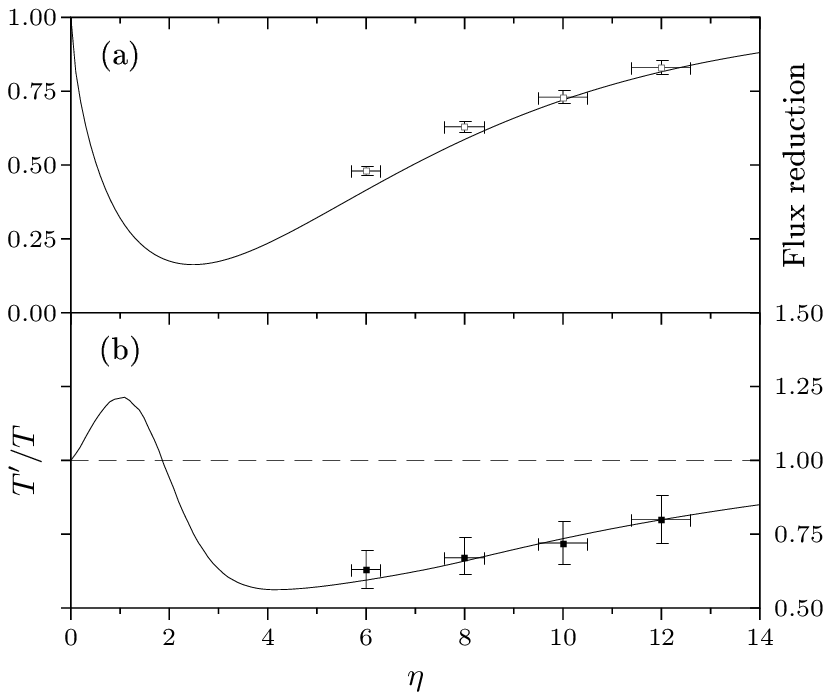}
\caption{(a):~Fraction of remaining atoms after the microwave horn
antenna as a function of $\eta$. The experimental points are in
good agreement with theory (solid line); the points at $\eta=6$
and 8 are slightly above the expected value, probably because of a
lack of microwave power to reach a 100\% efficiency in the
evaporation. (b):~Temperature change induced in the beam by
evaporation with a microwave at the parameter $\eta$, followed by
rethermalization. The solid line is the theoretical curve; the
points are experimental results of evaporation performed in the
interesting region where the phase-space density increases
significantly. For $\eta=8$, the phase-space density of the beam
increases by 2.6. For those experiments, the mean velocity is
$\bar{v}=0.8$~m/s and the guide gradient $600$~G/cm.}
\label{Fig:EvapMW}
\end{figure}

We calculate the fraction $\varphi(\nu_{\rm mw})$ of atoms
remaining after such a microwave antenna operating at $\nu_{\rm
mw}$, in the same way as for the single-radius evaporation scheme.
Defining $\eta$ as $h(\nu_{\rm hf}-\nu_{\rm mw})/(\kb T)$, we work
out the following approximate formula, valid for a linear
potential: $\varphi(\eta)=1-0.89\, \eta^{0.67}{\rm
e}^{-0.27\eta}$. The shape of this curve plotted on Fig.
\ref{Fig:EvapMW}~(a) is qualitatively similar to the one obtained
with RF evaporation, but with a much larger width and a more
pronounced minimum.

In practice, microwave evaporation is implemented by using a 10~dB
gain horn antenna powered by a microwave synthesizer delivering
typically 20~dBm after amplification. The directivity of such an
antenna allows us to use the microwave evaporation scheme even in
locations where nearby metallic walls prevent the use of RF
evaporation. For instance we can use it in the metal cross located
in the middle of the guide, by shining the microwave through
optical viewports (see Fig.~\ref{Fig:Setup}). We have used this
evaporation scheme in order to measure the beam temperature and
found a very good agreement (less than 10\% difference) between
microwave and RF spectroscopic measurements of temperature.

We have also calculated the temperature change $T'/T$ experienced
by the beam upon evaporation with a microwave antenna followed by
rethermalization. Fig. \ref{Fig:EvapMW}~(b) shows the
corresponding theoretical curve, and the experimental data points
obtained for evaporation with $\eta=6,\,8,\,10$ and~12. The beam
temperature was measured with a RF antenna. The agreement is
excellent, the theoretical curve having no adjustable parameters.
For the point corresponding to $\eta=8$, the gain in phase-space
density is $2.6$, which is significantly higher than the maximum
gain of $1.9$ that one can obtain with (single-radius) RF
evaporation.

\subsection{Ten-fold increase of the beam phase-space density with several antennas}
\label{Par:Gain10}

In order to achieve a significant gain in phase-space density, we
use the following setup (Fig. \ref{Fig:Setup}): We choose $v_{\rm
inj}=90$~cm/s and $b=800$~G/cm. Five RF antennas, separated by
$\sim20$~cm, are placed between $0.5\leqslant z\leqslant 1.3$~m
with frequencies decreasing from $49$ to 25~MHz, then the horn
antenna at $z=1.7$~m is operated at $\nu_{\rm mw}=6774$~MHz, and
finally five more RF antennas are placed at $2\leqslant z\leqslant
2.8$~m with frequencies decreasing from $21$ to $14$~MHz. We use
such a scheme, in which obviously no complete rethermalization can
occur between successive RF antennas\footnote{Even at the low
velocity achieved in the second, horizontal part of the guide, on
average only $\sim1.5$ collisions per atom can occur over 20~cm.},
because it is less sensitive to a potential lack of efficiency of
an antenna for some frequency. In addition, this scheme is closer
to the usual evaporation ramp in the time domain used for
evaporative cooling of trapped atoms than the step-like
evaporation process that would consist in letting the beam
rethermalize completely after each antenna.

The last evaporation antenna being placed at $z=2.8$~m, and the
analysis antenna used for the temperature measurements at $z=4$~m,
the beam can rethermalize completely after the last antenna, and
therefore we measure the temperature of a beam which is really in
thermal equilibrium.

\begin{figure}[t]
\includegraphics[width=7.5cm]{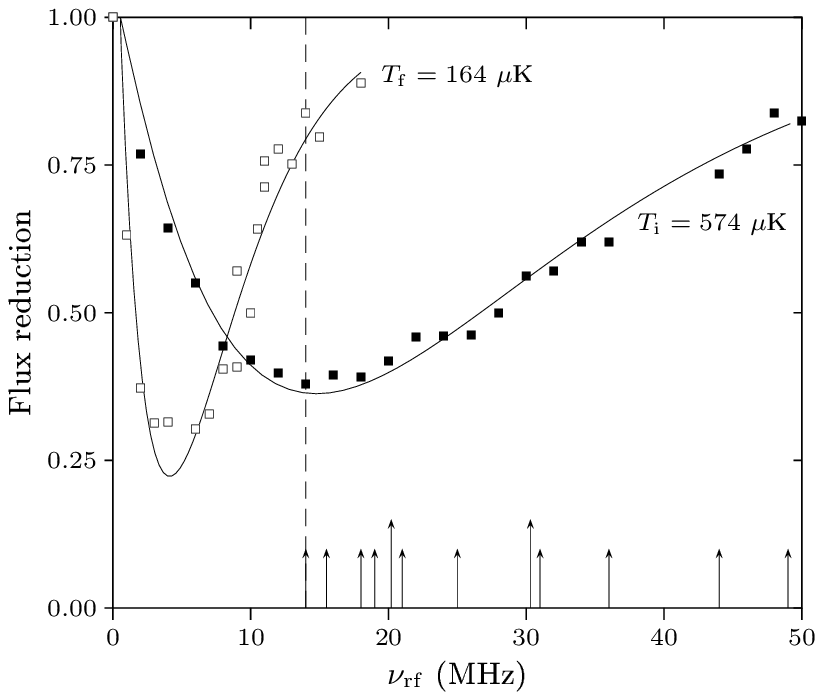}
\caption{Radio-frequency spectra used to determine the temperature
of the atomic beam with and without evaporation by 10 RF antennas
and the horn antenna. The temperature reduction from 574 to
164~$\mu$K, accompanied by a flux reduction by a factor 7.7, leads
to a ten-fold increase of the phase-space density. The arrows
represent the evaporation frequencies of the 10 RF antennas. The
two longer arrows correspond to RF frequencies that would lead to
the existence of evaporation radii coinciding with the ones
resulting from microwave evaporation by the horn; another one at
60.7~MHz is not shown on the figure.} \label{Fig:Gain10}
\end{figure}

Fig. \ref{Fig:Gain10} shows the RF spectra from which we deduce
the initial (without evaporation) and final (after evaporation by
the 10 RF and microwave antennas) temperatures $T_{\rm i}$ and
$T_{\rm f}$. We find $T_{\rm i}=574\pm 9$~$\mu$K, and $T_{\rm
f}=164\pm 6$~$\mu$K. The corresponding flux reduction is
$\Phi_{\rm f}/\Phi_{\rm i}=0.13\pm0.02$. From this we infer a gain
in phase-space density by a factor $10.4^{+4.1}_{-3.0}$. The
spectrum giving $T_{\rm f}$ corresponds to a well thermalized beam
(the spectrum does not reach the value 1 for, \emph{e.g.},
$\nu_{\rm rf}=14$~MHz, which would be the signature~\cite{PRLRb2}
of the absence of rethermalization after the last evaporation
antenna). This configuration corresponds to the existence of a
steady-state temperature gradient $dT/dz\simeq 200$~$\mu$K/m along
the beam.

Those measurements require a good stability of the incident atomic
flux. Indeed, a small relative variation $\delta \Phi_{\rm
i}/\Phi_{\rm i}$ can be amplified very significantly since a
decrease in $\Phi_{\rm _i}$ will result in less collisions between
successive antennas, and thus in a more pronounced reduction of
the flux than in the collisional case. We observed that
fluctuations in $\Phi_{\rm i}$ by 30\% could lead to fluctuations
above 50\% in $\Phi_{\rm f}$. For the data of
Fig.~\ref{Fig:Gain10} we therefore checked that the absolute flux
was almost constant and equal to its maximal value during the
whole data acquisition.

\section{Conclusion and prospects}

In this article, we have demonstrated the first significant gain
in the phase-space density of a magnetically guided atomic beam by
means of continuous evaporative cooling. This ten-fold increase
brings the beam's phase-space density to about~$2\times 10^{-7}$.
In order to gain the seven remaining orders of magnitude required
to reach quantum degeneracy, the temperature needs to be decreased
so much that the guide confinement will become harmonic. As
pointed out in \cite{epjd}, the evaporation becomes less efficient
for this confinement as compared to the linear case, and the
elastic collision rate cannot increase significantly during the
evaporation process.

In order to have a quantitative picture of the whole evaporation
ramp all the way to quantum degeneracy, we therefore calculated
for arbitrary values of $\alpha$, the change in flux and
temperature for a given value of $\eta$. This allows one to
calculate, through Eq. (\ref{Eq:Cr}) and (\ref{Eq:Psd}), the gain
in phase-space density and the evolution of the collision rate.

\begin{figure}
\includegraphics[width=8.5cm]{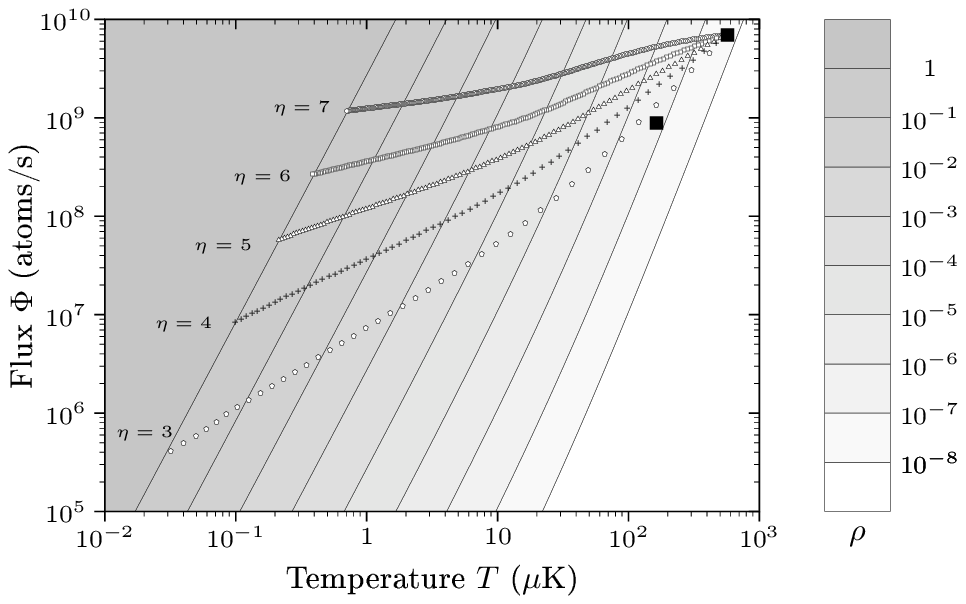}
\caption{Theoretical ``trajectories'' to quantum degeneracy
followed by the beam in the plane $(T,\,\Phi)$ during evaporation
with many antennas, at constant $\eta$. The grey scale represents
the on-axis phase-space density $\rho$ (which scales as
$\Phi/T^{7/2}$ in the linear regime $\alpha\ll 1$, and as
$\Phi/T^{5/2}$ in the harmonic regime). The atomic beam propagates
with a velocity of $60$~cm/s in a magnetic guide having a
transverse gradient of 800~G/cm, with a longitudinal bias field of
1~G. The initial flux is $7\times 10^9$ atoms/s, and the initial
temperature 570~$\mu$K. Each symbol in the various evaporation
trajectories corresponds to the action of one antenna. For higher
evaporation parameters $\eta$, the final flux and temperatures
(when $\rho=1$) are higher, but this requires more antennas (the
number of antennas required to reach $\rho=1$ for $\eta=3$, 4, 5,
6 and 7, is 41, 64, 106, 181, and 306, respectively). The black
squares represent the experimental beam parameters before and
after evaporation with the method described in
\S~\ref{Par:Gain10}.} \label{Fig:EvapPsd}
\end{figure}

\begin{figure}
\includegraphics[width=8.5cm]{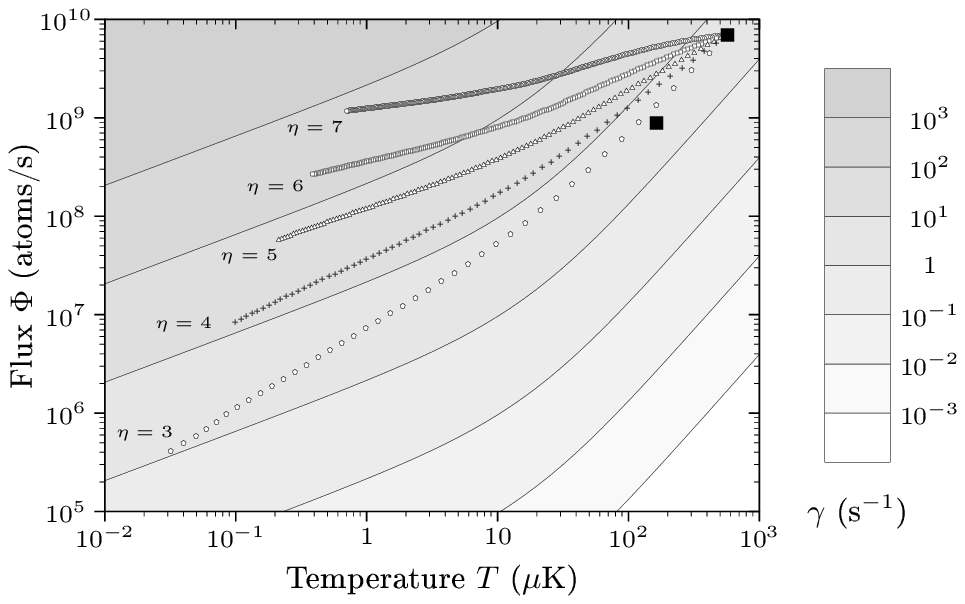}
\caption{Same as Fig \ref{Fig:EvapPsd}, but with a grey scale
representing the elastic collision rate $\gamma$, assuming that
the collision cross section is purely s-wave, and constant in
magnitude (both assumptions fail above typically 100~$\mu$K). The
cross-over from a linear trap (in which $\gamma\sim\Phi/T^{3/2}$)
to a harmonic trap (where $\gamma\sim\Phi/T^{1/2}$) is clearly
visible when $k_{\rm B}T$ becomes on the order of $\mu B_0$ (here,
at $T\sim 30$~$\mu$K since $B_0=1$~G). \emph{Runaway} evaporation,
\emph{ie} constant increase of $\gamma$, can only occur for large
values of $\eta$; for $\eta=4$, which requires a moderate number
of antennas, $\gamma$ starts by increasing at the beginning of the
evaporation, when the trap is still linear, but then decreases as
the guide becomes more and more harmonic.} \label{Fig:EvapColl}
\end{figure}

The resulting ``evaporation trajectories'' followed by the beam in
the plane $(T,\,\Phi)$ when it crosses successive single-radius
evaporation antennas are plotted on Fig.~\ref{Fig:EvapPsd}
and~\ref{Fig:EvapColl}. The frequency of each antenna is adapted
so as to evaporate at constant $\eta$ parameter, and we assume
that full rethermalization occurs after each antenna. One sees
that evaporating at relatively low $\eta$ minimizes the number of
antennas needed to reach degeneracy, but implies to have very low
final flux and temperatures, in the 30~nK range. Simultaneously,
the collision rate for such low evaporation parameters is
decreasing significantly in the final evaporation stages when the
temperature is low and the potential harmonic (see
Fig.~\ref{Fig:EvapColl}). On the contrary, evaporating at too
large $\eta$ requires too many antennas and thus a very long
guide.

A promising strategy probably consists in operating the antennas
with $\eta=4$ or 5, requiring a number of antennas between 60 and
100. Nonetheless, due to the finite range of the antennas, with
such a large number of evaporation zones one probably needs to
abandon the simple model of discrete evaporation zones followed by
rethermalization~\cite{mewes}, and resort to a continuous
evaporation model as the one used in standard BEC experiments and
considered in~\cite{mandonnet}. The atomic energy distribution is
then at any time close to a truncated Boltzmann
distribution~\cite{walraven}, whereas we considered here that
enough time was allowed between two successive antennas to recover
an exact equilibrium. Another possibility would consist in using
\emph{surface evaporation}~\cite{cornellJLTP} by approaching solid
surfaces close to the beam, on which the most energetic atoms
would stick: this would allow for the use of many very localized
evaporation zones.

\begin{acknowledgements}
We thank G.~Santarelli and P.~Bouyer for lending us microwave
synthesizers. Z.~Wang acknowledges support from the European Marie
Curie Grant MIF1-CT-2004-509423. G.~Reinaudi acknowledges support
from the D\'el\'egation G\'en\'erale de l'Armement (DGA). This
work was supported by the CNRS, the Ecole Normale Sup\'erieure,
and the D\'el\'egation G\'en\'erale de l'Armement.
\end{acknowledgements}

\end{document}